\documentclass[12pt,a4paper]{article}
\usepackage{graphicx}
\usepackage{authblk}
\usepackage{amsfonts}
\usepackage[dvips]{color}

\def\beq{\begin{equation}}
\def\eeq{\end{equation}}
\def\beqa{\begin{eqnarray}}
\def\eeqa{\end{eqnarray}}

\def\ket#1{|\,#1\,\rangle}

\def\braket#1#2{\langle\, #1\,|\,#2\,\rangle}

\def\red#1{\textcolor{red}{#1}}

\begin{document}

\title{The correspondence between mutually unbiased bases and mutually orthogonal extraordinary supersquares}

\author[1]{Iulia Ghiu\thanks{email: iulia.ghiu@g.unibuc.ro}}
\author[2]{Cristian Ghiu}
\affil[1]{{\small Centre for Advanced Quantum Physics,
Department of Physics, University of Bucharest,
P. O.  Box MG-11, R-077125  Bucharest-M\u{a}gurele, Romania}}
\affil[2]{{\small Department of Mathematics II, Faculty of Applied Sciences, University Politehnica of Bucharest, Romania}}

\maketitle

\begin{abstract}
We study the connection between mutually unbiased bases and mutually orthogonal extraordinary supersquares, a wider class of squares which does not contain only the Latin squares. We show that there are four types of complete sets of mutually orthogonal extraordinary supersquares for the dimension $d=8$. We introduce the concept of physical striation and show that this is equivalent to the extraordinary supersquare. The general algorithm for obtaining the mutually unbiased bases and the physical striations is constructed and it is shown that the complete set of mutually unbiased physical striations is equivalent to the complete set of  mutually orthogonal extraordinary supersquares. We apply the algorithm to two examples: one for two-qubit systems ($d=4$) and one for three-qubit systems ($d=8$), by using the Type II complete sets of mutually orthogonal extraordinary supersquares of order 8.
\end{abstract}


\section{Introduction}
The mutually unbiased bases are widely used in many protocols of quantum information processing: quantum tomography \cite{Wootters-1989,Klimov-2008,Manko-2011}, quantum cryptography \cite{cripto}, discrete Wigner function \cite{Wootters-1987,Gibbons}, understanding the complementarity \cite{Petz-2007,Petz-2009,Petz-2010,Petz-2012}, quantum teleportation \cite{telep}, quantum error correction codes \cite{Gottesman}, or the mean king's problem \cite{Hayashi}, just to enumerate a few of them.
Two bases $\{\ket{\psi_j}\}$ and $\{\ket{\phi_k}\}$ are called mutually unbiased if for any two vectors, one has $|\braket{\psi_j}{\phi_k}|^2=1/d$, where $d$ is the dimension of the Hilbert space \cite{Wootters-1989}.
The maximal number of mutually unbiased bases (MUBs) can be at most $d+1$ \cite{Ivanovic} and this value can be reached in the case when the dimension of the space is a prime or a power of prime \cite{Cal}.

MUBs can be constructed using different methods. One method consists in obtaining $d + 1$ classes of $d - 1$ commuting operators, whose eigenvectors represent the MUBs. These special operators are called mutually unbiased operators \cite{Band}.

Wootters \cite{Wootters-ibm} and Gibbons {\it et al.} \cite{Gibbons} proposed a geometrical approach that is based on the correspondence of the MUBs to the so-called discrete phase space. The phase space of a $d$-level system (qudit) is a $d\times d$ lattice, whose coordinates are the elements of the finite Galois field $\mathbb{F}_d$. A state is associated to a line in the discrete phase space and the set of parallel lines is called a striation \cite{Wootters-ibm}. The line passing through the origin is called ray. It turns out that the MUBs are determined by the bases associated with each striation.
A detailed review by Durt {\it et al.} presents different constructions of MUBs as well as their applications \cite{Durt-2010}.

Recently, the connection between magic and Latin squares with quantum information theory has been investigated. The quantum game based on a magic square allows two observers Alice and Bob to share an entangled quantum state \cite{Gawron,Pawela}. Also an interesting application of Latin squares is quantum teleportation, where generalized Bell states are given in terms of Latin squares \cite{Tanaka}.

Similarities between MUBs and mutually orthogonal Latin squares have been analyzed during the last years \cite{Wootters-ibm,Beth,Wootters-Found,Paterek-2009,Paterek-2010,Hall-C,Hall-2010,Rao-2010,Hall-phd,Klapp-2003}.
In Refs. \cite{Wootters-ibm,Wootters-Found}, one considers the case when the two striations with vertical and horizontal lines are present among the set of mutually orthogonal striations. This fact leads only to
a special class of MUBs, the one which contains the eigenvectors of tensor products of the Pauli operators $X$  and the identity, $Z$ and the identity, or their combinations. The associated striations of MUBs are Latin squares. In Ref. \cite{Paterek-2009}, a set of Latin operators is constructed, whose eigenvectors form a complete set of MUBs, therefore the associated striations are Latin squares. We should mention that the connection is done only with the Latin squares (also the squares with vertical and horizontal lines are considered) and not with a wider class.

There are at least two different ways of defining the problem of construction of MUBs:

A) the generation of the MU operators. This ends the problem and no connection to Latin squares is done.

B) the generation of both the MU operators and the striations associated to each basis. There are discussions on the link between striations and Latin squares. The striations are relevant if we want to compute the discrete Wigner function, since we need to know the expression of each line of all the striations.

In this paper we analyze the problem B), by trying to answer to three questions. Suppose that one starts with the ray denoted by the number 1 as below in the square of Fig. \ref{exemplu-intrebare}:
\begin{figure}[h]
\begin{center}
\begin{tabular}{|c|c|c|c|}
\hline    ?  & ? & ? &  {\bf 1} \\
\hline  ? & ? & ${\bf 1}$ & ? \\
\hline  ? & ${\bf 1}$& ? & ? \\
\hline  ${\bf 1}$ & ? & ? & ? \\
\hline
\end{tabular}
\end{center}
\caption{What properties should fulfill a striation associated to MUB?}
\label{exemplu-intrebare}
\end{figure}

(i) What properties should fulfill the ray such that the square in Fig. \ref{exemplu-intrebare} to be a striation?

(ii)  How can one fill in a Latin square in order to obtain a striation corresponding to an MUB?

(iii) Is it compulsory that a striation to be associated to a Latin square?

Our answers are based on a mathematical concept called extraordinary supersquare, which was recently introduced by us in Ref. \cite{Ghiu-matem}.

The paper is organized as follows. In Sec. 2 we briefly review the definitions of supersquares and extraordinary squares, as well as the construction of the mutually extraordinary supersquares of order 4. In Sec. 3 we obtain four types of mutually extraordinary supersquares of order 8 and present an example of complete set of Type II mutually extraordinary supersquares. In Sec. 4.1 we give the definition of the physical striation and further a theorem that shows the equivalence between the physical striation and the extraordinary supersquare. Sec. 4.2 presents the general algorithm for the construction of the MUBs and the physical striations for $d=p^n$, with $p$ prime, which consists in five steps. We give an example of complete set of MUBs for two qubits and an example for three qubits, each example being characterized by physical striations which are not all Latin squares or striations with vertical and horizontal lines. Finally, we make some concluding remarks in Sec. 5, by pointing out the answers to the three questions (i)$-$(iii) raised in the Introduction. For completeness, some basic concepts of finite fields $\mathbb{F}_{p^n}$ are reviewed in Appendix A.

\section{Preliminaries: Extraordinary supersquares}

\subsection{Definition of the supersquares and the construction of the mutually orthogonal supersquares}
A square of order $d$, denoted by $M=[M_{ij}]$ (with $i, j$ = 1, ..., $d$), is a $d\times d$ array from the numbers $ 1,...,d $ such that each number occurs $d$ times.
A Latin square of order $d$, denoted by $M=[M_{ij}]$ (with $i, j$ = 1, ..., $d$), is a $d\times d$ array from the numbers $ 1,...,d $ such that each number occurs in every row and every column only once. A row-Latin square of order $d$ is a square, where each row is a permutation of the $d$ numbers \cite{Mullen-1998}. A column-Latin square of order $d$ is a square, where each column is a permutation of the $d$ numbers. Two squares $M$ and $N$ are called orthogonal if all the pairs $(M_{ij}, N_{ij})$ are distinct.

If $d$ is a power of a prime number, then an equivalent definition of a square is to assign to each number in the square an element $(x_1,x_2)$ in the space $\mathbb{F}_d\times \mathbb{F}_d$, $\mathbb{F}_d$ being the finite field with $d$ elements. By $x_1$ we label the row and by $x_2$ the column of the square.

{\bf Definition 1.} \cite{Ghiu-matem} {\bf A square of order $d$} (with $d$ power of a prime number) is a partition of $\mathbb{F}_d\times \mathbb{F}_d$, being denoted as $S=\{ A_1, A_2,..., A_d\}$, where the number of elements of the subset $A_j$ is $d$.
To the elements of the subset $A_j$ we assign the number $j$:
\beqa
&&\mbox{the}\; \mbox{elements}\; \mbox{of} \; A_1 \longrightarrow 1\nonumber \\
&&\mbox{the}\; \mbox{elements}\; \mbox{of} \; A_2 \longrightarrow 2\label{nr-asociat} \\
&& ...\nonumber \\
&&\mbox{the}\; \mbox{elements}\; \mbox{of} \; A_d \longrightarrow d. \nonumber
\eeqa

In the case $d=4$, the elements of $\mathbb{F}_4$ are: $\{ 0, 1, \mu , \mu^2 \}$, $\mu $ being a primitive element of $\mathbb{F}_4$. We show in Fig. \ref{Lat-ex} the definition of a Latin square of order 4 as a partition.

\begin{figure}[h]
%\begin{figure*}[!ht]
\begin{center}
\begin{tabular}{c||c|c|c|c||}
\hline \hline $ \mu^2$ & 4  & 3  & 2 & 1\\
\hline $\mu$   & 3 & 4 & 1 & 2\\
\hline $1$     & 2 & 1 & 4 & 3\\
\hline $0$     & 1 & 2 & 3 & 4 \\
\hline   \hline      &$0$&$1$&$\mu$&$\mu^2$\\
\end{tabular}
\end{center}
\caption{ A Latin square of order 4 described as a partition of $\mathbb{F}_4\times \mathbb{F}_4$: $A_1=\{(0,0)$, $(1,1)$, $(\mu ,\mu )$, $(\mu^2,\mu^2)\}$; $A_2=\{(0,1)$, $(1,0)$, $(\mu ,\mu^2)$, $(\mu^2,\mu )\}$; $A_3=\{(0,\mu )$, $(1,\mu^2)$, $(\mu ,0)$, $(\mu^2,1)\}$; $A_4=\{(0,\mu^2)$, $(1,\mu )$, $(\mu ,1)$, $(\mu^2,0)\}$.}
\label{Lat-ex}
\end{figure}

The concept of a partition of a set is equivalent to that of an equivalence relation on that set \cite{Lipschutz}. In the case of a square, the equivalence classes are $A_1, A_2,..., A_d$.

{\bf Definition 2.} \cite{Ghiu-matem} Consider a square of order $d$ denoted by $S=\{A_1$, $A_2$, ..., $A_d\}$. $S$ is called a {\bf supersquare of order d} if $A_1$ is a subgroup with $d$ elements of $\mathbb{F}_d\times \mathbb{F}_d$ and $S=\mathbb{F}_d\times \mathbb{F}_d/A_1$ is the quotient set.

The subsets $A_j$ are as follows:
\beqa
&& A_1 = \hat{0}; \nonumber \\
&& A_2= a_2+A_1 = \hat a_2; \nonumber \\
&& .... \label{def-super} \\
&& A_d= a_d+A_1 = \hat a_d,\nonumber
\eeqa
where $a_j\in \mathbb{F}_d\times \mathbb{F}_d$, $j$= 2, 3, ..., $d$.  $A_1$ is called the generating subgroup of the supersquare.

The Latin square given in Fig. \ref{Lat-ex} is a supersquare: the generating subgroup is $A_1$=$\{(0,0)$,$(1,1)$,$(\mu ,\mu )$,$(\mu^2 ,\mu^2)\}$, while $A_2=(0,1)+A_1$, $A_3=(0,\mu )+A_1$, and $A_4=(0,\mu^2)+A_1$.

We proved that the maximum number of mutually orthogonal supersquares is $d+1$, this set being called complete \cite{Ghiu-matem}.
The construction of the complete set of mutually orthogonal supersquares consists of two steps \cite{Ghiu-matem}:

- step 1: the construction of the generating subgroups of each supersquare $A_1$, $B_1$,..., $Z_1$;

- step 2: the construction of the other $d-1$ equivalence classes of each supersquare, which do not contain the element zero.

\subsection{Definition of the extraordinary squares}

Let us consider $v_1=(x_1,y_1)$ and $v_2=(x_2,y_2)$ $\in \mathbb{F}_d\times \mathbb{F}_d$ with $d = p^n$, $p$ being a prime number. We denote by $\vert v_1\hspace{0.3cm} v_2 \vert$ the following determinant:
$$ \vert v_1\hspace{0.3cm} v_2 \vert = \left\vert
\begin{array}{cc}
x_1 & x_2\\
y_1 & y_2
\end{array}
\right\vert . $$

The trace of an element $\alpha \in \mathbb{F}_{p^n}$ is given in the Appendix. We denote by $K$ the subgroup of $\mathbb{F}_{p^n}$, whose elements have the trace equal to zero:
\beq
K=\{ \alpha \in \mathbb{F}_{p^n} : \hspace{0.1cm} \mbox{tr} \; \alpha =0\}.
\label{k}
\eeq

{\bf Definition 3.} \cite{Ghiu-matem} The subgroup $G\in \mathbb{F}_{p^n}\times \mathbb{F} _{p^n}$ is called {\bf extraordinary} if for any of its two elements $g_1$ and $g_2 \in G$, one has $\vert g_1 \hspace{0.3cm} g_2\vert \in K$.

{\bf Definition 4.} A square of order $d$, $S=\{A_1,...,A_d\}$ is called {\bf extraordinary} if there is $j\in \{1,2,...,d\}$ such that $A_j$ is an extraordinary subgroup of $\mathbb{F}_d\times \mathbb{F}_d$.

As a consequence, a supersquare is extraordinary if its generating subgroup is extraordinary.
The classification of the squares of order $d$ is shown below:
\begin{itemize}
\item Square: $S=A_1\cup A_2...\cup A_d$, with $A_j$ mutually disjoint subsets, each having $d$ elements.
\item Supersquare: $A_1$ is a subgroup and $A_j= a_j+A_1$, with $j=2, 3,..., d$.
\item Extraordinary squares: $A_1$ is an extraordinary subgroup.
\item Extraordinary supersquares: $A_1$ is an extraordinary subgroup and $A_j= a_j+A_1$, with $j=2, 3,..., d$.
\end{itemize}

\subsection{The complete sets of mutually orthogonal extraordinary supersquares of order 4}

In the case $d=4$, the complete set of mutually orthogonal extraordinary supersquares consists of five squares. Let us denote by  $A_1, B_1, C_1, D_1,$ and $E_1$ the generating extraordinary subgroups. We proved that for $d=4$, there are only two kinds of complete sets of  mutually orthogonal extraordinary supersquares \cite{Ghiu-matem}:

Type I) Let $\{ v_1,v_2\}$ be an arbitrary basis in $\mathbb{F}_4 \times \mathbb{F}_4$. The five generating extraordinary subgroups are:
$A_1= \mathbb{F}_4 \, v_1$; $B_1=\mathbb{F}_4\, v_2$; $C_1=\mathbb{F}_4\, (v_1+ \mu \, v_2)$;
$D_1=\mathbb{F}_4\, (v_1+ \mu ^2\, v_2)$;
$E_1=\mathbb{F}_4\, (v_1+ v_2)$.

Type II) Consider $v_1$ and $v_2\in \mathbb{F}_4 \times \mathbb{F}_4$ such that $ \vert v_1\hspace{0.3cm} v_2 \vert = 1$. Then the five generating extraordinary subgroups are as follows: $A_1= \mathbb{F}_4 \, v_1$;
$B_1=\mathbb{Z}_2\, v_2 + \mathbb{Z}_2\, (v_1+\mu \, v_2) $; $C_1=\mathbb{Z}_2\, \mu \, v_2 + \mathbb{Z}_2\, (\mu^2\, v_1+\mu^2 \, v_2)$; $D_1=\mathbb{Z}_2\, \mu^2\, v_2 + \mathbb{Z}_2\, (\mu \, v_1+\mu \, v_2) $;
$E_1=\mathbb{Z}_2\, (v_1+v_2) + \mathbb{Z}_2\, (\mu \, v_1+\mu^2 \, v_2)$.

An example of complete set of mutually orthogonal extraordinary supersquares of order 4 of Type II is shown in Fig. \ref{ex-nu-axa}, being obtained for $v_1=(1,\mu^2)$ and $v_2=(1,\mu )$. Only the square A) is Latin.

\begin{figure*}[!ht]
\begin{center}
\begin{tabular}{|c|c|c|c|}
\hline   4  &  \red{{\bf 1}} & 3 & 2 \\
\hline  3 & 2 & 4 & \red{{\bf 1}} \\
\hline  2 & 3 & \red{{\bf 1}} & 4  \\
\hline  \red{{\bf 1}} & 4 & 2 & 3  \\
\hline
\end{tabular}
\hspace{0.2cm}
\begin{tabular}{|c|c|c|c|}
\hline   3  &  2  &  2  &  3 \\
\hline  4 & \red{{\bf 1}} & \red{{\bf 1}} & 4 \\
\hline  2 & 3 & 3 & 2  \\
\hline  \red{{\bf 1}} & 4 & 4 & \red{{\bf 1}}  \\
\hline
\end{tabular}
\hspace{0.2cm}
\begin{tabular}{|c|c|c|c|}
\hline  $ \red{{\bf 1}} $ & $ 2 $ & $ \red{{\bf 1}} $ & $ 2  $\\
\hline  3 & 4 & 3 & 4 \\
\hline  3 & 4 & 3 & 4  \\
\hline  \red{{\bf 1}} & 2 & \red{{\bf 1}} & 2  \\
\hline
\end{tabular}
\hspace{0.2cm}
\begin{tabular}{|c|c|c|c|}
\hline  $ 4 $ & $ 2 $ & $ 3 $ & $ \red{{\bf 1}} $\\
\hline  \red{{\bf 1}} & 3 & 2 & 4 \\
\hline  4 & 2 & 3 & \red{{\bf 1}}  \\
\hline  \red{{\bf 1}} & 3 & 2 & 4  \\
\hline
\end{tabular}
\hspace{0.2cm}
\begin{tabular}{|c|c|c|c|}
\hline  $ 2 $ & $ 2 $ & $ 4 $ & $ 4 $\\
\hline  2 & 2 & 4 & 4 \\
\hline  \red{{\bf 1}} & \red{{\bf 1}} & 3 & 3  \\
\hline  \red{{\bf 1}} & \red{{\bf 1}} & 3 & 3  \\
\hline
\end{tabular}
\vspace{0.2cm}\\
A) \hspace{2.3cm} B) \hspace{2.3cm} C) \hspace{2.3cm} D) \hspace{2.3cm} E)
\end{center}
\caption{The complete set of mutually orthogonal extraordinary supersquares of order 4 of Type II. The elements which generate the supersquares are $v_1=(1,\mu ^2)$ and $v_2=(1,\mu )$ and we have: A) Latin, B) column-Latin, C) square, D) row-Latin, E) square. The generating extraordinary subgroup of each square is denoted by bold red 1.}
\label{ex-nu-axa}
\end{figure*}

\section{The construction of the complete sets of mutually orthogonal extraordinary supersquares of order 8}

The subgroup $K$ defined by Eq. (\ref{k}) for $d=8$ is $K=\{ 0, \mu , \mu^2, \mu^4\}$, $\mu $ being a primitive element (for details see Appendix A). Suppose that $G\subseteq \mathbb{F}_8\times \mathbb{F}_8$ is a subgroup which contains eight elements.
Let $v_1$, $v_2\in G$ with $\vert v_1\hspace{0.3cm} v_2 \vert = k$, where $k\in K$ and $k\ne 0$. Then $G$ is an extraordinary subgroup if and only if:
\beqa
&& \mbox{i)}\; G = \mathbb{F}_8 \, u \hspace{0.2cm} (u\ne 0,\; u\in \mathbb{F}_8\times \mathbb{F}_8) \hspace{0.3cm} \mbox{or}
\nonumber \\
&& \mbox{ii)}\; G = \mathbb{Z}_2\, v_1 + (Kk^{-1})\; v_2. \nonumber
\eeqa

According to Proposition 3.3 and Corollary 3.6 in Ref. \cite{Ghiu-matem}, in order to obtain the complete set of $d+1$ mutually orthogonal supersquares, it is sufficient to obtain the $d+1$ generating subgroups $A_1^{(1)}$, $A_1^{(2)}$,..., $A_1^{(d)}$ such that the intersection of any two of these subgroups is the element zero. We found four types of such 9 generating extraordinary subgroups in the case $d=8$, which are given below.  Therefore we obtain four different types of complete sets of mutually orthogonal extraordinary supersquares of order 8. In order to obtain the whole squares, one needs to construct the other $d-1$ equivalence classes of each supersquare \cite{Ghiu-matem}.

Type I) Let $\{ v_1,v_2\}$ be an arbitrary basis in $\mathbb{F}_8 \times \mathbb{F}_8$. The generating extraordinary subgroups are: $\mathbb{F}_d(v_1+\lambda_j\, v_2)$, $j$ = 0, 1,..., $d-1$ and $\mathbb{F}_dv_2$, where $\lambda_j\in \mathbb{F}_d$ are all distinct.

Further we consider $\{ v_1,v_2\}$ to be an arbitrary basis in $\mathbb{F}_8 \times \mathbb{F}_8$ such that $k=\vert v_1 \hspace{0.3cm} v_2\vert \in K$. We denote by $\tilde K=K\, k^{-1}=\{0, 1, k, k^3\}$.
The generating extraordinary subgroups of the other three types are as follows:

Type II) $A_1=\mathbb{Z}_2(v_2+k^4v_1)+\tilde Kv_1$; $B_1=\mathbb{Z}_2k^2v_1+\tilde K(k^5v_2+k^2v_1)$; $C_1=\mathbb{Z}_2k^4v_1+\tilde K(k^3v_2+k^6v_1)$; $D_1=\mathbb{Z}_2k^5v_1+\tilde K(k^2v_2+k^4v_1)$; $E_1=\mathbb{Z}_2k^6v_1+\tilde K(kv_2+v_1)$; $F_1=\mathbb{Z}_2(kv_1+kv_2)+\tilde Kk^6v_2$; $G_1=\mathbb{Z}_2kv_2+\tilde K(k^6v_1+k^6v_2)$; $H_1=\mathbb{Z}_2k^4v_2+\tilde K(k^3v_1+k^5v_2)$; $I_1=\mathbb{Z}_2(k^2v_1+k^3v_2)+\tilde K(v_1+k^6v_2)$.

Type III) $A_1=\mathbb{F}_8v_2$; $B_1=\mathbb{F}_8(v_1+v_2)$; $C_1=\mathbb{F}_8(kv_1+v_2)$; $D_1=\mathbb{Z}_2(v_2+k^2v_1)+\tilde Kv_1$; $E_1=\mathbb{Z}_2k^2v_1+\tilde K(k^5v_2+k^4v_1)$; $F_1=\mathbb{Z}_2k^4v_1+\tilde K(k^3v_2+k^5v_1)$; $G_1=\mathbb{Z}_2k^5v_1+\tilde K(k^2v_2+v_1)$; $H_1=\mathbb{Z}_2k^6v_1+\tilde K(kv_2+k^4v_1)$; $I_1=\mathbb{Z}_2(v_1+k^5v_2)+\tilde K(k^5v_1+kv_2)$.

Type IV) $A_1=\mathbb{F}_8v_2$; $B_1=\mathbb{Z}_2(v_2+k^2v_1)+\tilde Kv_1$; $C_1=\mathbb{Z}_2k^2v_1+\tilde K(k^5v_2+v_1)$; $D_1=\mathbb{Z}_2k^4v_1+\tilde K(k^3v_2+v_1)$; $E_1=\mathbb{Z}_2k^5v_1+\tilde K(k^2v_2+v_1)$; $F_1=\mathbb{Z}_2k^6v_1+\tilde K(kv_2+v_1)$; $G_1=\mathbb{Z}_2(k^2v_1+k^6v_2)+\tilde K(v_1+v_2)$; $H_1=\mathbb{Z}_2(k^2v_1+k^2v_2)+\tilde K(v_1+k^4v_2)$; $I_1=\mathbb{Z}_2(k^5v_1+k^5v_2)+\tilde K(v_1+k^6v_2)$.

\begin{figure*}[!ht]
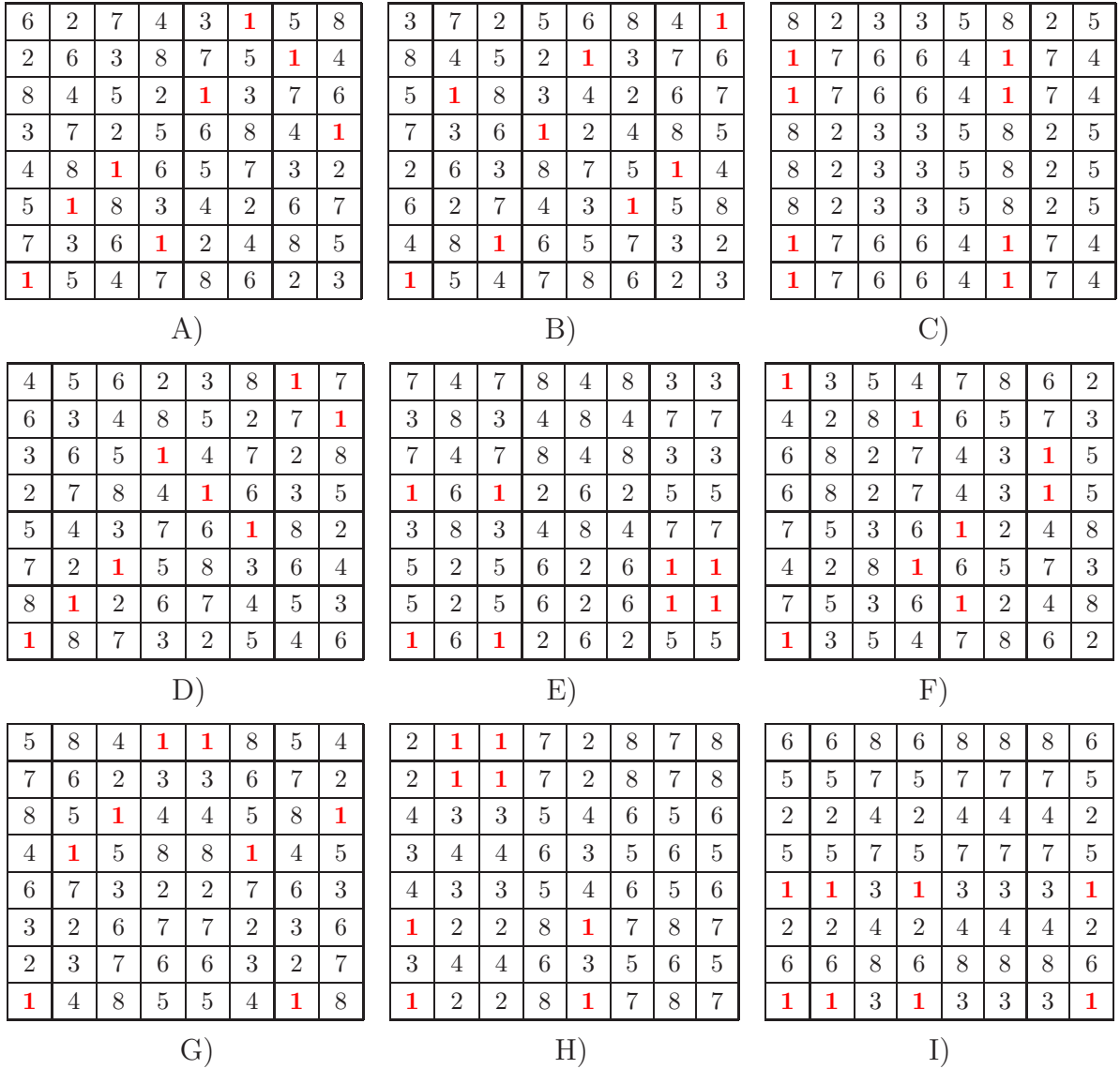

\begin{center}
\begin{tabular}{|c|c|c|c|c|c|c|c|}
\hline  {\footnotesize 6} & {\footnotesize 2} & {\footnotesize 7} & {\footnotesize 4} & {\footnotesize 3} & {\footnotesize \red{{\bf 1}}} & {\footnotesize 5} & {\footnotesize 8} \\
\hline  {\footnotesize 2} & {\footnotesize 6} & {\footnotesize 3} & {\footnotesize 8} & {\footnotesize 7} & {\footnotesize 5} & {\footnotesize \red{{\bf 1}}} & {\footnotesize 4}\\
\hline  {\footnotesize 8} & {\footnotesize 4} & {\footnotesize 5} & {\footnotesize 2} & {\footnotesize \red{{\bf 1}}} & {\footnotesize 3} & {\footnotesize 7} & {\footnotesize 6} \\
\hline  {\footnotesize 3} & {\footnotesize 7} & {\footnotesize 2} & {\footnotesize 5} & {\footnotesize 6} & {\footnotesize 8} & {\footnotesize 4} & {\footnotesize \red{{\bf 1}}} \\
\hline  {\footnotesize 4} & {\footnotesize 8} & {\footnotesize \red{{\bf 1}}} & {\footnotesize 6} & {\footnotesize 5} & {\footnotesize 7} & {\footnotesize 3} & {\footnotesize 2} \\
\hline  {\footnotesize 5} & {\footnotesize \red{{\bf 1}}} & {\footnotesize 8} & {\footnotesize 3} & {\footnotesize 4} & {\footnotesize 2} & {\footnotesize 6} & {\footnotesize 7} \\
\hline  {\footnotesize 7} & {\footnotesize 3} & {\footnotesize 6} & {\footnotesize \red{{\bf 1}}} & {\footnotesize 2} & {\footnotesize 4} & {\footnotesize 8} & {\footnotesize 5} \\
\hline  {\footnotesize \red{{\bf 1}}} & {\footnotesize 5} & {\footnotesize 4} & {\footnotesize 7} & {\footnotesize 8} & {\footnotesize 6} & {\footnotesize 2} & {\footnotesize 3} \\
\hline
\end{tabular}
\hspace{0.1cm}
\begin{tabular}{|c|c|c|c|c|c|c|c|}
\hline  {\footnotesize 3} & {\footnotesize 7} & {\footnotesize 2} & {\footnotesize 5} & {\footnotesize 6} & {\footnotesize 8} & {\footnotesize 4} & {\footnotesize \red{{\bf 1}}} \\
\hline  {\footnotesize 8} & {\footnotesize 4} & {\footnotesize 5} & {\footnotesize 2} & {\footnotesize \red{{\bf 1}}} & {\footnotesize 3} & {\footnotesize 7} & {\footnotesize 6} \\
\hline  {\footnotesize 5} & {\footnotesize \red{{\bf 1}}} & {\footnotesize 8} & {\footnotesize 3} & {\footnotesize 4} & {\footnotesize 2} & {\footnotesize 6} & {\footnotesize 7} \\
\hline  {\footnotesize 7} & {\footnotesize 3} & {\footnotesize 6} & {\footnotesize \red{{\bf 1}}} & {\footnotesize 2} & {\footnotesize 4} & {\footnotesize 8} & {\footnotesize 5} \\
\hline  {\footnotesize 2} & {\footnotesize 6} & {\footnotesize 3} & {\footnotesize 8} & {\footnotesize 7} & {\footnotesize 5} & {\footnotesize \red{{\bf 1}}} & {\footnotesize 4} \\
\hline  {\footnotesize 6} & {\footnotesize 2} & {\footnotesize 7} & {\footnotesize 4} & {\footnotesize 3} & {\footnotesize \red{{\bf 1}}} & {\footnotesize 5} & {\footnotesize 8} \\
\hline {\footnotesize 4} & {\footnotesize 8} & {\footnotesize \red{{\bf 1}}} & {\footnotesize 6} & {\footnotesize 5} & {\footnotesize 7} & {\footnotesize 3} & {\footnotesize 2} \\
\hline  {\footnotesize \red{{\bf 1}}} & {\footnotesize 5} & {\footnotesize 4} & {\footnotesize 7} & {\footnotesize 8} & {\footnotesize 6} & {\footnotesize 2} & {\footnotesize 3} \\
\hline
\end{tabular}
\hspace{0.1cm}
\begin{tabular}{|c|c|c|c|c|c|c|c|}
\hline  {\footnotesize 8} & {\footnotesize 2} & {\footnotesize 3} & {\footnotesize 3} & {\footnotesize 5} & {\footnotesize 8} & {\footnotesize 2} & {\footnotesize 5} \\
\hline  {\footnotesize \red{{\bf 1}}} & {\footnotesize 7} & {\footnotesize 6} & {\footnotesize 6} & {\footnotesize 4} & {\footnotesize \red{{\bf 1}}} & {\footnotesize 7} & {\footnotesize 4} \\
\hline {\footnotesize \red{{\bf 1}}} & {\footnotesize 7} & {\footnotesize 6} & {\footnotesize 6} & {\footnotesize 4} & {\footnotesize \red{{\bf 1}}} & {\footnotesize 7} & {\footnotesize 4} \\
\hline  {\footnotesize 8} & {\footnotesize 2} & {\footnotesize 3} & {\footnotesize 3} & {\footnotesize 5} & {\footnotesize 8} & {\footnotesize 2} & {\footnotesize 5} \\
\hline  {\footnotesize 8} & {\footnotesize 2} & {\footnotesize 3} & {\footnotesize 3} & {\footnotesize 5} & {\footnotesize 8} & {\footnotesize 2} & {\footnotesize 5} \\
\hline  {\footnotesize 8} & {\footnotesize 2} & {\footnotesize 3} & {\footnotesize 3} & {\footnotesize 5} & {\footnotesize 8} & {\footnotesize 2} & {\footnotesize 5} \\
\hline  {\footnotesize \red{{\bf 1}}} & {\footnotesize 7} & {\footnotesize 6} & {\footnotesize 6} & {\footnotesize 4} & {\footnotesize \red{{\bf 1}}} & {\footnotesize 7} & {\footnotesize 4} \\
\hline  {\footnotesize \red{{\bf 1}}} & {\footnotesize 7} & {\footnotesize 6} & {\footnotesize 6} & {\footnotesize 4} & {\footnotesize \red{{\bf 1}}} & {\footnotesize 7} & {\footnotesize 4} \\
\hline
\end{tabular}
\vspace{0.2cm}\\
A) \hspace{4.5cm} B) \hspace{4.5cm} C)
\vspace{0.2cm}\\
\begin{tabular}{|c|c|c|c|c|c|c|c|}
\hline  {\footnotesize 4} & {\footnotesize 5} & {\footnotesize 6} & {\footnotesize 2} & {\footnotesize 3} & {\footnotesize 8} & {\footnotesize \red{{\bf 1}}} & {\footnotesize 7} \\
\hline  {\footnotesize 6} & {\footnotesize 3} & {\footnotesize 4} & {\footnotesize 8} & {\footnotesize 5} & {\footnotesize 2} & {\footnotesize 7} & {\footnotesize \red{{\bf 1}}} \\
\hline  {\footnotesize 3} & {\footnotesize 6} & {\footnotesize 5} & {\footnotesize \red{{\bf 1}}} & {\footnotesize 4} & {\footnotesize 7} & {\footnotesize 2} & {\footnotesize 8} \\
\hline  {\footnotesize 2} & {\footnotesize 7} & {\footnotesize 8} & {\footnotesize 4} & {\footnotesize \red{{\bf 1}}} & {\footnotesize 6} & {\footnotesize 3} & {\footnotesize 5} \\
\hline  {\footnotesize 5} & {\footnotesize 4} & {\footnotesize 3} & {\footnotesize 7} & {\footnotesize 6} & {\footnotesize \red{{\bf 1}}} & {\footnotesize 8} & {\footnotesize 2} \\
\hline  {\footnotesize 7} & {\footnotesize 2} & {\footnotesize \red{{\bf 1}}} & {\footnotesize 5} & {\footnotesize 8} & {\footnotesize 3} & {\footnotesize 6} & {\footnotesize 4} \\
\hline  {\footnotesize 8} & {\footnotesize \red{{\bf 1}}} & {\footnotesize 2} & {\footnotesize 6} & {\footnotesize 7} & {\footnotesize 4} & {\footnotesize 5} & {\footnotesize 3} \\
\hline  {\footnotesize \red{{\bf 1}}} & {\footnotesize 8} & {\footnotesize 7} & {\footnotesize 3} & {\footnotesize 2} & {\footnotesize 5} & {\footnotesize 4} & {\footnotesize 6} \\
\hline
\end{tabular}
\hspace{0.1cm}
\begin{tabular}{|c|c|c|c|c|c|c|c|}
\hline  {\footnotesize 7} & {\footnotesize 4} & {\footnotesize 7} & {\footnotesize 8} & {\footnotesize 4} & {\footnotesize 8} & {\footnotesize 3} & {\footnotesize 3} \\
\hline  {\footnotesize 3} & {\footnotesize 8} & {\footnotesize 3} & {\footnotesize 4} & {\footnotesize 8} & {\footnotesize 4} & {\footnotesize 7} & {\footnotesize 7} \\
\hline  {\footnotesize 7} & {\footnotesize 4} & {\footnotesize 7} & {\footnotesize 8} & {\footnotesize 4} & {\footnotesize 8} & {\footnotesize 3} & {\footnotesize 3} \\
\hline  {\footnotesize \red{{\bf 1}}} & {\footnotesize 6} & {\footnotesize \red{{\bf 1}}} & {\footnotesize 2} & {\footnotesize 6} & {\footnotesize 2} & {\footnotesize 5} & {\footnotesize 5} \\
\hline  {\footnotesize 3} & {\footnotesize 8} & {\footnotesize 3} & {\footnotesize 4} & {\footnotesize 8} & {\footnotesize 4} & {\footnotesize 7} & {\footnotesize 7} \\
\hline  {\footnotesize 5} & {\footnotesize 2} & {\footnotesize 5} & {\footnotesize 6} & {\footnotesize 2} & {\footnotesize 6} & {\footnotesize \red{{\bf 1}}} & {\footnotesize \red{{\bf 1}}} \\
\hline  {\footnotesize 5} & {\footnotesize 2} & {\footnotesize 5} & {\footnotesize 6} & {\footnotesize 2} & {\footnotesize 6} & {\footnotesize \red{{\bf 1}}} & {\footnotesize \red{{\bf 1}}} \\
\hline  {\footnotesize \red{{\bf 1}}} & {\footnotesize 6} & {\footnotesize \red{{\bf 1}}} & {\footnotesize 2} & {\footnotesize 6} & {\footnotesize 2} & {\footnotesize 5} & {\footnotesize 5} \\
\hline
\end{tabular}
\hspace{0.1cm}
\begin{tabular}{|c|c|c|c|c|c|c|c|}
\hline  {\footnotesize \red{{\bf 1}}} & {\footnotesize 3} & {\footnotesize 5} & {\footnotesize 4} & {\footnotesize 7} & {\footnotesize 8} & {\footnotesize 6} & {\footnotesize 2} \\
\hline  {\footnotesize 4} & {\footnotesize 2} & {\footnotesize 8} & {\footnotesize \red{{\bf 1}}} & {\footnotesize 6} & {\footnotesize 5} & {\footnotesize 7} & {\footnotesize 3} \\
\hline  {\footnotesize 6} & {\footnotesize 8} & {\footnotesize 2} & {\footnotesize 7} & {\footnotesize 4} & {\footnotesize 3} & {\footnotesize \red{{\bf 1}}} & {\footnotesize 5} \\
\hline  {\footnotesize 6} & {\footnotesize 8} & {\footnotesize 2} & {\footnotesize 7} & {\footnotesize 4} & {\footnotesize 3} & {\footnotesize \red{{\bf 1}}} & {\footnotesize 5} \\
\hline  {\footnotesize 7} & {\footnotesize 5} & {\footnotesize 3} & {\footnotesize 6} & {\footnotesize \red{{\bf 1}}} & {\footnotesize 2} & {\footnotesize 4} & {\footnotesize 8} \\
\hline  {\footnotesize 4} & {\footnotesize 2} & {\footnotesize 8} & {\footnotesize \red{{\bf 1}}} & {\footnotesize 6} & {\footnotesize 5} & {\footnotesize 7} & {\footnotesize 3} \\
\hline  {\footnotesize 7} & {\footnotesize 5} & {\footnotesize 3} & {\footnotesize 6} & {\footnotesize \red{{\bf 1}}} & {\footnotesize 2} & {\footnotesize 4} & {\footnotesize 8} \\
\hline  {\footnotesize \red{{\bf 1}}} & {\footnotesize 3} & {\footnotesize 5} & {\footnotesize 4} & {\footnotesize 7} & {\footnotesize 8} & {\footnotesize 6} & {\footnotesize 2} \\
\hline
\end{tabular}
\vspace{0.2cm}\\
D) \hspace{4.5cm} E) \hspace{4.5cm} F)
\vspace{0.2cm}\\
\begin{tabular}{|c|c|c|c|c|c|c|c|}
\hline  {\footnotesize 5} & {\footnotesize 8} & {\footnotesize 4} & {\footnotesize \red{{\bf 1}}} & {\footnotesize \red{{\bf 1}}} & {\footnotesize 8} & {\footnotesize 5} & {\footnotesize 4} \\
\hline  {\footnotesize 7} & {\footnotesize 6} & {\footnotesize 2} & {\footnotesize 3} & {\footnotesize 3} & {\footnotesize 6} & {\footnotesize 7} & {\footnotesize 2} \\
\hline  {\footnotesize 8} & {\footnotesize 5} & {\footnotesize \red{{\bf 1}}} & {\footnotesize 4} & {\footnotesize 4} & {\footnotesize 5} & {\footnotesize 8} & {\footnotesize \red{{\bf 1}}} \\
\hline  {\footnotesize 4} & {\footnotesize \red{{\bf 1}}} & {\footnotesize 5} & {\footnotesize 8} & {\footnotesize 8} & {\footnotesize \red{{\bf 1}}} & {\footnotesize 4} & {\footnotesize 5} \\
\hline  {\footnotesize 6} & {\footnotesize 7} & {\footnotesize 3} & {\footnotesize 2} & {\footnotesize 2} & {\footnotesize 7} & {\footnotesize 6} & {\footnotesize 3} \\
\hline  {\footnotesize 3} & {\footnotesize 2} & {\footnotesize 6} & {\footnotesize 7} & {\footnotesize 7} & {\footnotesize 2} & {\footnotesize 3} & {\footnotesize 6} \\
\hline  {\footnotesize 2} & {\footnotesize 3} & {\footnotesize 7} & {\footnotesize 6} & {\footnotesize 6} & {\footnotesize 3} & {\footnotesize 2} & {\footnotesize 7} \\
\hline  {\footnotesize \red{{\bf 1}}} & {\footnotesize 4} & {\footnotesize 8} & {\footnotesize 5} & {\footnotesize 5} & {\footnotesize 4} & {\footnotesize \red{{\bf 1}}} & {\footnotesize 8} \\
\hline
\end{tabular}
\hspace{0.1cm}
\begin{tabular}{|c|c|c|c|c|c|c|c|}
\hline   {\footnotesize 2} & {\footnotesize \red{{\bf 1}}} & {\footnotesize \red{{\bf 1}}} & {\footnotesize 7} & {\footnotesize 2} & {\footnotesize 8} & {\footnotesize 7} & {\footnotesize 8} \\
\hline  {\footnotesize 2} & {\footnotesize \red{{\bf 1}}} & {\footnotesize \red{{\bf 1}}} & {\footnotesize 7} & {\footnotesize 2} & {\footnotesize 8} & {\footnotesize 7} & {\footnotesize 8} \\
\hline  {\footnotesize 4} & {\footnotesize 3} & {\footnotesize 3} & {\footnotesize 5} & {\footnotesize 4} & {\footnotesize 6} & {\footnotesize 5} & {\footnotesize 6} \\
\hline  {\footnotesize 3} & {\footnotesize 4} & {\footnotesize 4} & {\footnotesize 6} & {\footnotesize 3} & {\footnotesize 5} & {\footnotesize 6} & {\footnotesize 5} \\
\hline  {\footnotesize 4} & {\footnotesize 3} & {\footnotesize 3} & {\footnotesize 5} & {\footnotesize 4} & {\footnotesize 6} & {\footnotesize 5} & {\footnotesize 6} \\
\hline  {\footnotesize \red{{\bf 1}}} & {\footnotesize 2} & {\footnotesize 2} & {\footnotesize 8} & {\footnotesize \red{{\bf 1}}} & {\footnotesize 7} & {\footnotesize 8} & {\footnotesize 7} \\
\hline  {\footnotesize 3} & {\footnotesize 4} & {\footnotesize 4} & {\footnotesize 6} & {\footnotesize 3} & {\footnotesize 5} & {\footnotesize 6} & {\footnotesize 5} \\
\hline  {\footnotesize \red{{\bf 1}}} & {\footnotesize 2} & {\footnotesize 2} & {\footnotesize 8} & {\footnotesize \red{{\bf 1}}} & {\footnotesize 7} & {\footnotesize 8} & {\footnotesize 7} \\
\hline
\end{tabular}
\hspace{0.1cm}
\begin{tabular}{|c|c|c|c|c|c|c|c|}
\hline  {\footnotesize 6} & {\footnotesize 6} & {\footnotesize 8} & {\footnotesize 6} & {\footnotesize 8} & {\footnotesize 8} & {\footnotesize 8} & {\footnotesize 6} \\
\hline  {\footnotesize 5} & {\footnotesize 5} & {\footnotesize 7} & {\footnotesize 5} & {\footnotesize 7} & {\footnotesize 7} & {\footnotesize 7} & {\footnotesize 5} \\
\hline  {\footnotesize 2} & {\footnotesize 2} & {\footnotesize 4} & {\footnotesize 2} & {\footnotesize 4} & {\footnotesize 4} & {\footnotesize 4} & {\footnotesize 2} \\
\hline  {\footnotesize 5} & {\footnotesize 5} & {\footnotesize 7} & {\footnotesize 5} & {\footnotesize 7} & {\footnotesize 7} & {\footnotesize 7} & {\footnotesize 5} \\
\hline  {\footnotesize \red{{\bf 1}}} & {\footnotesize \red{{\bf 1}}} & {\footnotesize 3} & {\footnotesize \red{{\bf 1}}} & {\footnotesize 3} & {\footnotesize 3} & {\footnotesize 3} & {\footnotesize \red{{\bf 1}}} \\
\hline  {\footnotesize 2} & {\footnotesize 2} & {\footnotesize 4} & {\footnotesize 2} & {\footnotesize 4} & {\footnotesize 4} & {\footnotesize 4} & {\footnotesize 2} \\
\hline  {\footnotesize 6} & {\footnotesize 6} & {\footnotesize 8} & {\footnotesize 6} & {\footnotesize 8} & {\footnotesize 8} & {\footnotesize 8} & {\footnotesize 6} \\
\hline  {\footnotesize \red{{\bf 1}}} & {\footnotesize \red{{\bf 1}}} & {\footnotesize 3} & {\footnotesize \red{{\bf 1}}} & {\footnotesize 3} & {\footnotesize 3} & {\footnotesize 3} & {\footnotesize \red{{\bf 1}}} \\
\hline
\end{tabular}
\vspace{0.2cm}\\
G) \hspace{4.5cm} H) \hspace{4.5cm} I)
\end{center}
\caption{An example of complete set of mutually orthogonal extraordinary supersquares of order 8 of Type II. The elements which generate the supersquares are $v_1=(1,\mu )$ and $v_2=(\mu^3,\mu^2 )$ and we have: A) Latin, B) Latin, C) square, D) Latin, E) square, F) row-Latin, G) column-Latin, H) square, I) square. The generating extraordinary subgroup of each square is denoted by bold red 1.}
\label{ex-8}
\end{figure*}

We present in Fig. \ref{ex-8} an example of complete set of mutually orthogonal extraordinary supersquares of Type II of order 8. The generating basis is $v_1=(1,\mu)$ and $v_2=(\mu^3,\mu^2)$.

\section{The construction of mutually unbiased bases}

\subsection{Definition of the physical striation and its equivalence to the extraordinary supersquare}

In order to define a discrete Wigner function for the dimension power of prime $d=p^n$, Wootters \cite{Wootters-ibm} and Gibbons {\it et al.} \cite{Gibbons} constructed the discrete phase space with the help of $d^2$ points $(x_1,x_2)$, where $x_1\in \mathbb{F}_d$ runs along the horizontal axis and $x_2\in \mathbb{F}_d$ along the vertical one. To a quantum state one has to assign a line, which is defined as a subset of $d$ points. The line which passes through the origin is called a ray \cite{Gibbons}. Two lines in the discrete phase space are called parallel if they do not intersect. The discrete phase space consists of $d$ parallel lines. A given set of $d$ parallel lines generates a striation \cite{Wootters-ibm,Gibbons}. Two striations are mutually unbiased if each line of the first striation has exactly one intersecting point with each line of the second striation \cite{Wootters-Found}. There are $d+1$ mutually unbiased striations. Using the concepts introduced in Sec. 2.1, Definition 1, a striation has to be described by a square of order $d$, i.e. a striation is a partition of $\mathbb{F}_d\times \mathbb{F}_d$. The line in the discrete phase space is analogue of a subset (equivalence class) $A_j$ of the square. The correspondence between an equivalence class (line) $A_j$ and a quantum state is made by the function $Q$ \cite{Gibbons}: $Q(A_j)$ is the projection operator of a pure state, where $j$ = 1, 2, ..., $d$. The equivalent concepts of the geometry of discrete phase space and squares are shown in Table \ref{notatii}.

\begin{table*}
\begin{center}
\begin{tabular}{|c|c|c|}
\hline \hline Notation & Discrete phase space & Square \\
\hline \hline
$(x_1,x_2)$ & point  & element of $\mathbb{F}_d\times \mathbb{F}_d$ \\
\hline $A_j$ & line &  equivalence class\\
\hline $\{ 0, a_1,..., a_{d-1}\}$& ray & extraordinary subgroup\\
\hline  $S$ & striation & extraordinary square \\
\hline  $S$ & physical striation & extraordinary supersquare\\
& & (according to the Theorem) \\
\hline \hline
\end{tabular}
\caption{Table of equivalent concepts}
\label{notatii}
\end{center}
\end{table*}

Let us consider two elements $x$, $y\in \mathbb{F}_{p^n}$ which can be written with the help of two bases $E = \{ e_1,..., e_n\} $ and $F = \{ f_1,..., f_n\} $ as follows:
$
x=\sum_{i=1}^nx_{e_i}\, e_i$;
$y=\sum_{j=1}^ny_{f_j}\, f_j.
$
We denote by $X, Y, Z$ the generalized Pauli operators. To the translation in the phase space $\mathbb{F}_{p^n}\times \mathbb{F}_{p^n}$ by the element $(x,y)$ we associate an operator $T_{(x,y)}$ called the translation operator \cite{Gibbons}:
\beq
T_{(x,y)}:= X^{x_{e_1}}\, Z^{y_{f_1}}\otimes ... \otimes X^{x_{e_n}}\, Z^{y_{f_n}}.
\label{coresp}
\eeq

Let us consider a striation to be the square $S=\{ A_1,A_2,..., A_d \}$. If $Q(A_j)$ is the projection operator associated to the equivalence class $A_j$ ($j$ = 1, 2,..., $d-1$), then one needs \cite{Gibbons}:
\beq
Q((x,y)+A_j)=T_{(x,y)}\, Q(A_j)\, T^\dagger _{(x,y)}.
\label{invarianta}
\eeq
This requirement is imposed in order that the discrete Wigner function to be translationally covariant.

Let us denote by $T_{a_1}, T_{a_2},..., T_{a_{d-1}}$ with $a_j\in \mathbb{F}_d\times \mathbb{F}_d$ the set of $d-1$ commuting translation operators, whose common eigenvectors are $\ket{w_1}$,..., $\ket{w_d}$.
A striation is associated to the common eigenvectors, i.e. to the basis $\{ \ket{w_1}$,..., $\ket{w_d}\}$. The ray of this striation is $\{ 0, a_1,..., a_{d-1}\}$. One knows that if
\beq
\mbox{tr}(x_1\, y_2)=\mbox{tr}(x_2\, y_1),
\label{cond-com}
\eeq
 then the two translation operators (\ref{coresp}) $T_{(x_1,y_1)}$ and $T_{(x_2,y_2)}$ commute \cite{Klimov-2007,Ghiu-2012,Ghiu-2013}. If we denote by $a_1=(x_1,y_1)$ and by $a_2=(x_2,y_2)\in \mathbb{F}_d\times \mathbb{F}_d$, then the condition (\ref{cond-com}) is equivalent to $\vert a_1 \hspace{0.3cm} a_2\vert \in K$, where $K$ is the subgroup of $\mathbb{F}_d$ defined by Eq. (\ref{k}). In order to determine the set of $d-1$ commuting operators $T_{a_1}, T_{a_2},..., T_{a_{d-1}}$ of a given striation, one needs to obtain the subgroup $\{ 0, a_1,..., a_{d-1}\}$ in $\mathbb{F}_d\times \mathbb{F}_d$ such that $\vert a_i \hspace{0.3cm} a_j\vert \in K$ for all $i\ne j$, i.e. the subgroup has to be extraordinary according to the Definition 3 of Sec. 2.2. Therefore, the square associated to a striation must be an extraordinary square, since all the $d-1$ translation operators corresponding to the subgroup $\{ 0, a_1,..., a_{d-1}\}$, i.e. the ray, mutually commute (see Table \ref{notatii}).

 Wootters pointed out that a striation is invariant under the translations by the vectors $a_1, a_2,..., a_{d-1}$ \cite{Wootters-ibm}, i.e. the nonzero elements of the ray (see the step no. 3 and the example presented above it in Sec. 5 of \cite{Wootters-ibm}). Using the fact that the striation must remain invariant under the translation by the nonzero elements of the ray \cite{Wootters-ibm}, we give the following definition:

{\bf Definition 5.} Let us consider an extraordinary subgroup $A_1= \{ 0, a_1,..., a_{d-1}\}$ in $\mathbb{F}_d\times \mathbb{F}_d$. The extraordinary square $S=\{ A_1, A_2,..., A_d\}$ is called {\bf a physical striation} if $a_k+A_j=A_j$, where $k$ = 1, 2,..., $d-1$ and $j$ = 1, 2,..., $d$.

In other words, the physical striation $S$ contains an extraordinary subgroup and it is invariant under the translation by the nonzero elements $a_k$ of the extraordinary subgroup.

{\bf Theorem.} {\it An extraordinary square is a physical striation if and only if it is a supersquare.
The complete set of $d+1$ mutually unbiased physical striations is equivalent to the complete set of $d+1$ mutually orthogonal extraordinary supersquares.}

{\it Proof.} Firstly we will prove that if the extraordinary square is a physical striation, then the square must be a supersquare. The extraordinary square is $S=\{ A_1, A_2,..., A_d\}$ with  $A_1= \{ 0, a_1,..., a_{d-1}\}$. Let us denote the subset $A_2$ as $A_2=\{ u_1$, $u_2$,..., $u_d\}$. Since $A_1$ and $A_2$ are disjoint subsets, it means that $u_i\ne a_k$. We have that $A_2+a_k=A_2$ for all $k$ = 1, 2,..., $d-1$, which for a fixed $i$ leads to
\beqa
&&u_i+a_1\in A_2\nonumber \\
&&u_i+a_2\in A_2\nonumber \\
&&...\nonumber \\
&&u_i+a_{d-1}\in A_2,\nonumber
\eeqa
This can be rewritten as $u_i+A_1\subseteq A_2$. Since the number of elements of $u_i+A_1$ is the same as the number of elements of $A_2$, i.e. $d$, one obtains that $A_2=u_i+A_1$, where $u_i$ is any arbitrary $u_1$, $u_2$,..., $u_d$. The same argumentation is used for the other subsets $A_j$ with $j$ = 3,..., $d$ and this means that $S$ is an extraordinary supersquare.

Secondly, we need to prove the converse statement: if the extraordinary square $S$ is a supersquare, then $S$ is a physical striation. Since $S$ is supersquare, its equivalence classes have the structure given by Eq. (\ref{def-super}). The invariance under translations by the nonzero elements of $A_1$ is obvious, because $A_1$ is a subgroup.

As a consequence, we obtain that the complete set of $d+1$ mutually unbiased physical striations is equivalent to the complete set of $d+1$ mutually orthogonal extraordinary supersquares.

\subsection{The general algorithm for the construction of the MUBs for $d=p^n$}

The general algorithm for the construction of the complete set of MUBs is as follows:

1. The complete set of $d+1$ mutually orthogonal extraordinary supersquares is obtained. This is an independent mathematical problem to be solved \cite{Ghiu-matem}. We want to emphasize that the definition of the extraordinary supersquares is a mathematical one and no connection with the quantum states or operators is made in this definition.

2. We write the $d+1$ generating subgroups of the extraordinary supersquares.

3. To each element of these subgroups we assign its associated translation operator according to Eq. (\ref{coresp}). Since the subgroups are extraordinary, we obtain that the $d-1$ operators corresponding to each subgroup commute, i.e. we have a set of mutually unbiased operators.

4. For each set of commuting operators, we derive its set of common eigenvectors. The $d+1$ families of common eigenvectors represent the set of MUBs.

5. Further, we have to find the unique correspondence between a certain state of a basis of the MUBs and a certain equivalence class of the square (line of the striation). For a given physical striation, we choose one of the common eigenvectors to correspond to the generating extraordinary subgroup $A_1$ of the supersquare (i.e. the ray). Then, since the other classes $A_k$ ($k$ = 2, 3, ..., $d$) are obtained as $A_k= a_k+A_1$, we have that the eigenvector $Q(A_k) = T_{a_k}\, Q(A_1)\, T^\dagger _{a_k}$ as we get from Eq. (\ref{invarianta}). In this way, the correspondence is well defined and unique \cite{Wootters-ibm}.

Our algorithm is a generalization of the algorithm of Wootters \cite{Wootters-ibm}. In the construction of Wootters, one assumes that the two striations with vertical and horizontal lines are always present and, accordingly, one generates only a special class of MUBs, the one which contains the eigenvectors of tensor products of the Pauli operators $X$  and the identity, $Z$ and the identity, or their combination. In our algorithm, it is not necessary to start the construction with the squares with vertical and horizontal lines. Instead, we emphasize that one has to obtain the set of mutually orthogonal extraordinary supersquares as a first step in our construction.

In the case of systems consisting of two qubits, i.e. $d=4$, all the complete sets of five mutually orthogonal extraordinary supersquares  were presented in Sec. 2.3, the generating extraordinary subgroups being of Type I or II \cite{Ghiu-matem}. For a three-qubit system, i.e. $d=8$, we obtained in Sec. 3 four kinds of complete sets of mutually orthogonal extraordinary supersquares: Type I $-$ IV.

\subsection{An example of complete set of MUBs for $d=4$}

We denote by $X, Y$, and $Z$ the Pauli operators. Using the selfdual basis $\{ \mu,\mu^2\}$ in $\mathbb{F}_4$, we obtain the translation operators (\ref{coresp}): $T_{(0,1)}= Z\otimes Z$; $T_{(0,\mu )}= Z\otimes I$; $T_{(1,0)}= X\otimes X$; $T_{(\mu ,0)}= X\otimes I$.

Further we give an example of the construction of MUBs for $d=4$. We follow the steps 1 $- $ 5 given in Sec. 4.2 in order to obtain the MUBs. The five mutually orthogonal extraordinary supersquares of order 4 of Type II were obtained in Fig. \ref{ex-nu-axa} and this represents step 1 in our construction.
Steps 2 and 3 are shown in the table below, i.e. the generating extraordinary subgroups and the MU operators:
\[
\begin{tabular}{|c|c|c|}
\hline \hline Label & Generating extraordinary & Mutually unbiased \\
& subgroups & operators \\ \hline
\hline  $A_1$ & $(1,\mu^2); (\mu ,1); (\mu^2,\mu ) $ & $X\otimes Y; Y\otimes Z; Z\otimes X $ \\
\hline  $B_1$ & $(1,\mu ); (\mu^2 ,0); (\mu ,\mu ) $ & $Y\otimes X; I\otimes X; Y\otimes I  $  \\
\hline  $C_1$ & $(\mu ,\mu^2 ); (0,\mu ^2); (\mu ,0) $  & $X\otimes Z; I\otimes Z; X\otimes I  $\\
\hline  $D_1$ & $(\mu^2,1); (\mu^2,\mu^2); (0,\mu ) $  &  $Z\otimes Y; I\otimes Y; Z\otimes I $\\
\hline  $E_1$ & $(0,1); (1,0); (1,1) $ & $Z\otimes Z; X\otimes X; Y\otimes Y $\\
\hline \hline
\end{tabular}
\]

Step 4 consists in obtaining the MUBs, which represent the set of common eigenvectors:

\[
\begin{tabular}{|c|c|}
\hline \hline No. & Mutually unbiased bases  \\ \hline
\hline  1. & $\frac{1}{2}\left(
\begin{array}{c}
-i \\
i \\
1 \\
1
\end{array}
\right) $; $\frac{1}{2}\left(
\begin{array}{c}
i \\
i \\
-1 \\
1
\end{array}
\right) $;
$\frac{1}{2}\left(
\begin{array}{c}
i \\
-i \\
1 \\
1
\end{array}
\right) $;
$\frac{1}{2}\left(
\begin{array}{c}
-i \\
-i \\
1 \\
-1
\end{array}
\right) $
 \\
\hline  2. & $\frac{1}{2}\left(
\begin{array}{c}
i \\
i \\
-1 \\
-1
\end{array}
\right) $; $\frac{1}{2}\left(
\begin{array}{c}
i \\
-i \\
1 \\
-1
\end{array}
\right) $;
$\frac{1}{2}\left(
\begin{array}{c}
i \\
-i \\
-1 \\
1
\end{array}
\right) $;
$\frac{1}{2}\left(
\begin{array}{c}
i \\
i \\
1 \\
1
\end{array}
\right) $ \\
\hline  3. & $\frac{1}{\sqrt 2}\left(
\begin{array}{c}
1 \\
0 \\
1 \\
0
\end{array}
\right) $; $\frac{1}{\sqrt 2}\left(
\begin{array}{c}
0 \\
i \\
0 \\
i
\end{array}
\right) $;
$\frac{1}{\sqrt 2}\left(
\begin{array}{c}
-i \\
0 \\
i \\
0
\end{array}
\right) $;
$\frac{1}{\sqrt 2}\left(
\begin{array}{c}
0 \\
1 \\
0 \\
-1
\end{array}
\right) $
 \\
\hline  4. & $\frac{1}{\sqrt 2}\left(
\begin{array}{c}
i \\
-1 \\
0 \\
0
\end{array}
\right) $; $\frac{1}{\sqrt 2}\left(
\begin{array}{c}
0 \\
0 \\
i \\
-1
\end{array}
\right) $;
$\frac{1}{\sqrt 2}\left(
\begin{array}{c}
0 \\
0 \\
-1 \\
i
\end{array}
\right) $;
$\frac{1}{\sqrt 2}\left(
\begin{array}{c}
-1 \\
i \\
0 \\
0
\end{array}
\right) $
 \\
\hline  5. & $\frac{1}{\sqrt 2}\left(
\begin{array}{c}
1 \\
0 \\
0 \\
1
\end{array}
\right) $; $\frac{1}{\sqrt 2}\left(
\begin{array}{c}
-i \\
0 \\
0 \\
i
\end{array}
\right) $;
$\frac{1}{\sqrt 2}\left(
\begin{array}{c}
0 \\
i \\
i \\
0
\end{array}
\right) $;
$\frac{1}{\sqrt 2}\left(
\begin{array}{c}
0 \\
1 \\
-1 \\
0
\end{array}
\right) $
 \\
\hline \hline
\end{tabular}
\]

\begin{figure*}[!ht]
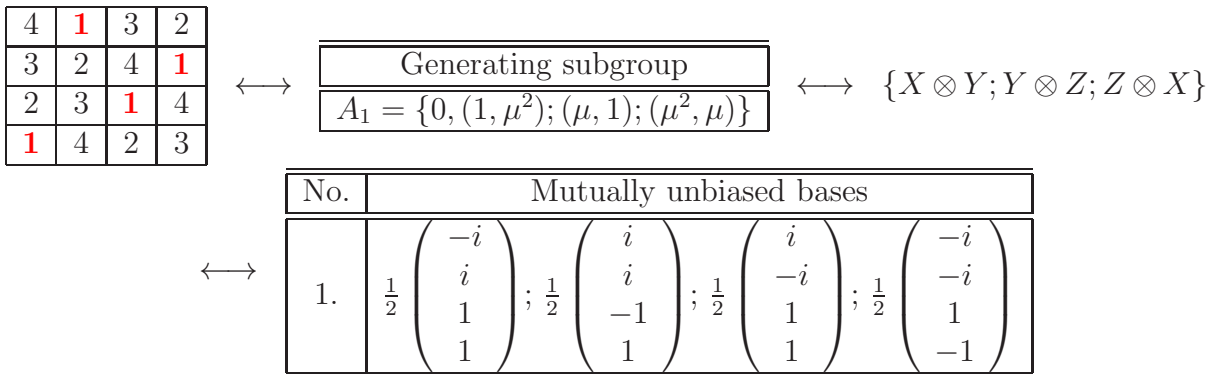

\begin{center}
\begin{tabular}{|c|c|c|c|}
\hline   4 & \red{{\bf 1}} & 3 & 2 \\
\hline  3 & 2 & 4 & \red{{\bf 1}} \\
\hline  2 & 3 & \red{{\bf 1}} & 4  \\
\hline  \red{{\bf 1}} & 4 & 2 & 3  \\
\hline
\end{tabular}
\hspace{0.1cm}
$\longleftrightarrow  $
\hspace{0.1cm}
\begin{tabular}{|c|}
\hline  \hline Generating subgroup \\ \hline
\hline  $A_1=\{ 0, (1,\mu^2); (\mu ,1); (\mu^2,\mu )\} $  \\
\hline
\end{tabular}
\hspace{0.1cm}
$\longleftrightarrow  $
\hspace{0.1cm}
$\{ X\otimes Y; Y\otimes Z; Z\otimes X\} $
\vspace{0.2cm}
\hspace{0.1cm} $\longleftrightarrow  $ \hspace{0.1cm}
\begin{tabular}{|c|c|}
\hline  \hline No. & Mutually unbiased bases  \\ \hline
\hline  1. & $\frac{1}{2}\left(
\begin{array}{c}
-i \\
i \\
1 \\
1
\end{array}
\right) $; $\frac{1}{2}\left(
\begin{array}{c}
i \\
i \\
-1 \\
1
\end{array}
\right) $;
$\frac{1}{2}\left(
\begin{array}{c}
i \\
-i \\
1 \\
1
\end{array}
\right) $;
$\frac{1}{2}\left(
\begin{array}{c}
-i \\
-i \\
1 \\
-1
\end{array}
\right) $
\\
\hline
\end{tabular}
\end{center}
\caption{The five steps for obtaining the MUBs described in Sec. 4.2. We discuss here only the first basis. In the step 5 we need to find the unique correspondence between an element of the basis and an equivalence class of the extraordinary supersquare. We associate the first vector $\ket{v_1}$ of the first basis to the extraordinary subgroup denoted by bold red 1 in the supersquare. Since the second equivalence class is $A_2=(1,\mu )+A_1$, we obtain the unique correspondence of the second vector of the basis $\ket{v_2}=T_{(1,\mu )}\ket{v_1} $ by using Eq. (\ref{invarianta}).}
\label{coresp-patr-baze}
\end{figure*}

The set of physical striations coincide with the set of mutually orthogonal extraordinary supersquares, according to the Theorem, and this is shown in Fig. \ref{ex-nu-axa}. The unique correspondence between extraordinary supersquares and MUBs is shown in Fig. \ref{coresp-patr-baze}. We present in the figure how the steps 1 $-$ 5 given in Sec. 4.2 are implemented.

\subsection{An example of complete set of MUBs for $d=8$ corresponding to the structure (0,9,0)}

One knows that any set of MUBs contains entangled bases \cite{Romeo}. In the case of three qubits, the MUBs are characterized by $(n_f,n_b,n_{ns})$, where $n_f$ is the number of completely factorized bases, $n_b$ is the number of biseparable bases that contain bipartite entanglement, while $n_{nf}$ is the number of nonseparable bases. The three numbers $(n_f,n_b,n_{ns})$ defines a structure of the set of MUBs. For three-qubit systems, there are four kinds of structures: (3,0,6), (1,6,2), (2,3,4), and (0,9,0) \cite{Romeo}.

The MUBs are useful tools for quantum tomography. One may ask which structure of MUBs for three qubits can be optimally implemented. The answer to this question is based on the concept of physical complexity of MUBs \cite{Klimov-2008}. The physical complexity is proportional to the total number of CNOT gates required for the tomography.
The measurements which involve entangled MUBs depend on nonlocal gate operations, which are very difficult to be performed with high accuracy. Therefore the optimal implementation of MUB tomography is based on that family of MUBs that requires a minimum number of nonlocal operations and this condition is satisfied by the structure (0,9,0) \cite{Klimov-2008}. In conclusion, among all the types of MUBs, the ones described by bipartite entanglement are of greatest importance in tomography. Below we present one example of MUBs with the structure (0,9,0).

Using the selfdual basis $\{ \mu^3,\mu^5, \mu^6 \}$ in $\mathbb{F}_8$, we obtain the translation operators (\ref{coresp}): $ T_{(0,\mu^3)}= Z\otimes I\otimes I$; $T_{(0,\mu^5 )}= I\otimes Z\otimes I$; $T_{(0,\mu^6 )}= I\otimes I\otimes Z$; $T_{(\mu^3,0)}= X\otimes I\otimes I$; $T_{(\mu^5 ,0)}= I\otimes X\otimes I$; $T_{(\mu^6 ,0)}= I\otimes I\otimes X$.

The first step of our algorithm is already accomplished in Sec. 3, where the complete set of Type II orthogonal extraordinary supersquares was obtained in the case when $v_1=(1,\mu )$ and $v_2=(\mu^3,\mu^2 )$, namely Fig. \ref{ex-8}. These represent also the physical striations.
Steps 2 and 3 consist in writing the eight generating extraordinary subgroups of the squares and assigning the mutually unbiased operators. The nine families of seven commuting operators are shown in Table \ref{operatori-8}. The complete set of MUBs generated by Table \ref{operatori-8} corresponds to the structure (0,9,0).

\begin{table*}
\begin{tabular}{|c|c|}
\hline \hline {\footnotesize Label of the} & {\footnotesize Mutually unbiased operators}\\
{\footnotesize generat. subgr.} & \\ \hline
\hline  {\footnotesize $A_1$} & {\footnotesize $X\otimes Y\otimes Y; Z\otimes Y\otimes X; Y\otimes I\otimes Z; Z\otimes I\otimes X; Y\otimes Y\otimes Z; I\otimes Y\otimes I; X\otimes I\otimes Y $} \\
\hline  {\footnotesize $B_1$} & {\footnotesize $X\otimes Z\otimes Y; Z\otimes Y\otimes I; Y\otimes X\otimes Y; Y\otimes X\otimes I; Z\otimes Y\otimes Y; X\otimes Z\otimes I; I\otimes I\otimes Y $} \\
\hline  {\footnotesize $C_1$} & {\footnotesize $Z\otimes I\otimes Z; I\otimes Z\otimes I; X\otimes Z\otimes X; Z\otimes Z\otimes Z; Y\otimes Z\otimes Y; X\otimes I\otimes X; Y\otimes I\otimes Y $} \\
\hline  {\footnotesize $D_1$} & {\footnotesize $Y\otimes Y\otimes Y; I\otimes Y\otimes Y; Y\otimes I\otimes I; I\otimes X\otimes Z; Y\otimes Z\otimes X; I\otimes Z\otimes X; Y\otimes X\otimes Z $} \\
\hline  {\footnotesize $E_1$} & {\footnotesize $Z\otimes Y\otimes Z; I\otimes Z\otimes Y; Z\otimes X\otimes X; Z\otimes Z\otimes Y; I\otimes X\otimes X; Z\otimes I\otimes I; I\otimes Y\otimes Z $} \\
\hline  {\footnotesize $F_1$} & {\footnotesize $X\otimes Y\otimes Z; Y\otimes Z\otimes I; Z\otimes X\otimes Z; X\otimes Y\otimes I; I\otimes I\otimes Z; Z\otimes X\otimes I; Y\otimes Z\otimes Z $} \\
\hline  {\footnotesize $G_1$} & {\footnotesize $Y\otimes X\otimes X; Z\otimes X\otimes Y; X\otimes I\otimes Z; Y\otimes I\otimes X; I\otimes X\otimes I; X\otimes X\otimes Z; Z\otimes I\otimes Y $} \\
\hline  {\footnotesize $H_1$} & {\footnotesize $X\otimes Y\otimes X; I\otimes X\otimes Y; X\otimes Z\otimes Z; X\otimes X\otimes Y; I\otimes Z\otimes Z; X\otimes I\otimes I; I\otimes Y\otimes X $} \\
\hline  {\footnotesize $I_1$} & {\footnotesize $I\otimes I\otimes X; X\otimes X\otimes X; X\otimes X\otimes I; Y\otimes Y\otimes X; Y\otimes Y\otimes I; Z\otimes Z\otimes I; Z\otimes Z\otimes X $} \\
\hline \hline
\end{tabular}
\caption{The mutually unbiased operators that correspond to the mutually orthogonal extraordinary supersquares of order 8 of Type II given in Fig. \ref{ex-8}. This is the structure (0,9,0).}
\label{operatori-8}
\end{table*}

\section{Conclusions}
In this paper we have investigated the connection between mutually unbiased bases and mutually orthogonal extraordinary supersquares. Firstly we have obtained four types of complete sets of mutually orthogonal extraordinary supersquares of order 8.

Secondly, we have introduced the concept of physical striation, which is based on the invariance of the extraordinary square associated to that striation under the translation by the nonzero elements of the ray. We have proved in the Theorem that a physical striation is equivalent to an extraordinary supersquare.

In conclusion, we answered to the questions raised in the Introduction. If we start with the striation of Fig. \ref{exemplu-intrebare}, then:
\begin{itemize}
\item the number '1' represents a subgroup of $\mathbb{F}_d\times \mathbb{F}_d $;
\item the subgroup described by '1' must be an extraordinary one;
\item the square has to be a supersquare;
\item it is not compulsory that the square to be a Latin one.
\end{itemize}
In other words, since a physical striation is described by an extraordinary supersquare, then the complete set of physical striations associated to the MUBs is given by the complete set of mutually orthogonal extraordinary supersquares.

We have proposed an algorithm for obtaining the complete set of MUBs, which consists in five steps. This algorithm was applied to an example for two qubits and an example for three qubits, which correspond to the structure (0,9,0). The fact that not only the Latin squares are accepted as physical striations may have strong implications in quantum information processes, e.g. quantum tomography.

\section*{Acknowledgements}
Iulia Ghiu thanks Gunnar Bj\"{o}rk, Andrei B. Klimov, and Luis L. S\'anchez-Soto for interesting discussions on mutually unbiased bases. The work of Iulia Ghiu was supported by CNCS - UEFISCDI, postdoctoral research project PD code 151, no. 150/30.07.2010 for the University of Bucharest.

\appendix
\section{Basic concepts of finite fields $\mathbb{F}_{p^n}$}

For $\alpha \in \mathbb{F}_{p^n}$, the trace is given by
$
\mbox{tr} \; \alpha = \alpha +\alpha ^p + \alpha ^{p^2}+ ... + \alpha^{p^{n-1}}.
$

Let us denote by $E = \{ e_1,..., e_n\} $  a basis in $\mathbb{F}_{p^n}$ and by $F = \{ f_1,..., f_n\} $ a second basis in $\mathbb{F}_{p^n}$. Then, $F$ is called the dual basis of $E$ if
$\mbox{tr}\, (e_i\, f_j) = \delta _{ij}.$ A basis $E$ is called selfdual if it is dual to itself.

In the case when $d=2^2$, there is only one irreducible polynomial: $P(x)=x^2+x+1$. Since $\mu $ is the primitive element of $\mathbb{F}_4$, we have $\mu^2 +\mu +1=0$. The elements of $\mathbb{F}_4$ are:
$\mathbb{F}_4=\{ 0, 1, \mu , \mu^2 \}$. The basis $\{ \mu,\mu^2\}$ is selfdual in $\mathbb{F}_4$.

In the case when $d=2^3$, we consider the irreducible polynomial: $P(x)=x^3+x+1$. Therefore the elements of $\mathbb{F}_8$ are:
$\mathbb{F}_8=\{ 0, 1, \mu , \mu^2 , \mu^3 , \mu^4 , \mu^5 , \mu^6 \}$. The basis $\{ \mu^3,\mu^5, \mu^6 \}$ in $\mathbb{F}_8$ is selfdual.

\end{document}